\documentclass{ws-ijqi}
\usepackage[english]{babel}
\usepackage{xcolor}
\usepackage{amsmath}
\usepackage{hyperref}

\newcommand{\ket}[1]{\vert {#1} \rangle}
\newcommand{\pure}[1]{\vert {#1} \rangle \langle {#1} \vert}

\newcommand{\beq}{\begin{equation}}
\newcommand{\eeq}{\end{equation}}

\begin{document}
\markboth{Nicola Dalla Pozza \and  Matteo G. A. Paris}
{An effective iterative method to build the Naimark extension of rank-n POVMs}
\title{AN EFFECTIVE ITERATIVE METHOD TO BUILD THE NAIMARK EXTENSION OF RANK-N POVMs}
\author{NICOLA DALLA POZZA}
\address{Quantum Technology Lab, 
Dipartimento di Fisica, Universit\`a di Milano, I-20133 
Milano, Italy \\ nicoladallapozza@gmail.com}
\author{MATTEO G. A. PARIS} 
\address{Quantum Technology Lab, 
Dipartimento di Fisica, Universit\`a di Milano, I-20133 
Milano, Italy \\ matteo.paris@fisica.unimi.it}
\maketitle
\begin{history}
\received{\today}
\end{history}
\catchline{}{}{}{}{}
\begin{abstract}
We revisit the problem of finding the Naimark extension of a probability 
operator-valued measure (POVM), i.e. its  implementation as a projective measurement in a larger Hilbert space. In particular, we suggest an 
iterative method to build the projective measurement from the sole 
requirements of orthogonality and positivity. Our method improves existing 
ones, as it may be employed also to extend POVMs containing elements with 
rank larger than one. It is also more effective in terms of computational 
steps.
\end{abstract}
\section{Introduction}
Any (generalised) measurement performed on  a physical system is described 
by a probability operator-valued measure (POVM) acting on the Hilbert space 
of the system. Naimark theorem  \cite{nmk40,akh63,hel73,hel76,hol01} ensures 
that any POVM may be implemented as a projective measurement in a larger 
Hilbert space, which is usually referred to as the {\em Naimark extension} 
of the POVM. As a matter of fact, there are infinite Naimark extensions 
and the theorem also ensures that a {\em canonical extension} exists, 
i.e. an implementation as an indirect measurement, where the system 
under investigation is coupled to an independently prepared probe 
system  \cite{per90} and then only the probe is subject to a (projective) measurement 
\cite{bin07,ber10,mtqm}.
\par

The problem of finding the Naimark extensions of a POVM is indeed a 
central one in quantum technology. On the one hand, it provides a 
concrete model to realize the measurement\cite{ban97,Zmeas}, and thus to 
assess entanglement cost\cite{ecos03} and/or implementations 
on different platforms \cite{tam1,ben07,ben10,lev89,spar14,fey16}. On  the other hand, 
it permits to evaluate the post-measurement state and thus to 
investigate the tradeoff between information gain and measurement 
disturbance \cite{igmd1,igmd2,igmd3,igmd4,igmd5,igmd6,wil13}, 
as well as any procedure aimed at quantum control \cite{iqc}.
\par
Let us consider a set of operators $\{\Pi_m\}$ that constitute a POVM 
for the physical system $S$ described by the Hilbert space 
$\mathcal{H}_S$, i.e.
\beq
\sum_{m=1}^M \Pi_m = I_S,\quad \Pi_m = \Pi_m^\dagger, 
\quad \Pi_m \geq 0\,.
\eeq
The elements of the set are not necessarily projectors,  
$\Pi_n \Pi_m \neq \Pi_n \delta_{n,m}$.
The Naimark theorem states that it is possible to {\em extend}  
each POVM elements to a larger (product) Hilbert space 
$\mathcal{H}_A \otimes \mathcal{H}_S$ (see \ref{a:kronecker})
such that the extended measurement operators are projectors in 
the product space.
In particular, it is possible to define the auxiliary Hilbert 
space $\mathcal{H}_A$ such that the system 
Hilbert space $\mathcal{H}_S$ is isomorphic to a subspace in $\mathcal{H}_A \otimes \mathcal{H}_S$, 
where the density operator $\rho$ defined on $\mathcal{H}_S$ corresponds to the 
density operator $\pure{e_1} \otimes \rho$ defined on $\mathcal{H}_A \otimes \mathcal{H}_S$. 
The state $\ket{e_1}$ may be chosen as the state corresponding 
to the first vector of the canonical basis of $\mathcal{H}_A$.
Naimark theorem states that we can find projectors $\{E_m\}$
\beq
E_m E_n = E_m \delta_{m,n}, \quad E_m = E_m^\dagger, \quad E_m \geq 0,
\label{vincoliEm}
\eeq
each of them corresponding to a POVM element $\Pi_m$ in the following 
sense. The distributions of the $m$-th outcome, as obtained from 
$\{E_m\}$ and $\{\Pi_m\}$ on the states $\pure{e_1} \otimes 
\rho$ and $\rho$ respectively, are the same, i.e. 
\beq
\hbox{Tr}_A\big[\Pi_m\,\rho\big] = \hbox{Tr}_{AS}\big[ 
\left(\pure{e_1} \otimes \rho\right)\, E_m\big]\,.
\label{BornRule}
\eeq
At the operatorial level, this is expressed by the following 
the set of relations
\beq
\Pi_m = \hbox{Tr}_A \left[ \left(\pure{e_1} \otimes 
{\mathbb I}_S\right)\, E_m\right]
\eeq
which, solved for the $E_m$ given the $\Pi_m$, provide the
desired Naimark extension of the POVM.
\par
As it was originally suggested by Helstrom\cite{hel76} the projectors 
$\{E_m\}$ may be built by placing a copy of $\Pi_m$ in the upper-left 
block position of the matrix representation of $\{E_m\}$ (corresponding 
to the element 1 in the matrix $e_1 \cdot e_1^T$). At the same time, no
explicit recipes had been provided on how to find the remaining blocks. 
The aim of this  paper is to describe an  iterative method for effectively
building those blocks upon exploiting the sole requirements of orthogonality 
and positivity. 
\par
The problem has been addressed before \cite{per95,pre98}, and constructive 
methods to find the projective measurement have been suggested. In short, 
these methods amount to set up and solve a linear problem which gives the 
coefficients of the projectors in the canonical basis of the enlarged Hilbert 
space $\mathcal{H}_A \otimes \mathcal{H}_S$. 
However, the focus has been on solving the problem for rank-1 POVM elements. 
Our iterative method, also based on solving a linear problem, 
shows two main advantages compared to existing techniques. 
On the one hand, it is more efficient in terms of computational steps 
and, on the other hand, it may applied also to POVMs containing 
elements with rank greater than one. 
\par
The paper is structured as follows. In the next Section we introduce 
the iterative method, first illustrating the basic idea and then, in 
Sections \ref{costruzioneProiettore} and 
\ref{costruzioneOrtogonale}, describing in details its two building 
blocks, i.e. the constrained building of an idempotent matrix and the 
constrained building of a matrix orthogonal to a given one. 
In Section \ref{ss:algo} we put everything together and illustrate 
the overall algorithm 
to build the Naimark extension of a generic rank-n POVM. In Section 
\ref{s:exa}, we illustrate few examples of application, whereas Section 
\ref{s:outro} closes the paper with some concluding remarks.
\section{An iterative method to build the Naimark extension of rank-n POVMs}
In the following, we will write  projectors as matrices of suitable 
sizes composed by blocks. 
The first step in building the projectors $E_m$  is analogue to the 
original Helstrom recipe, 
i.e. we define the upper-left block in the matrix of $E_m$ equal 
to $\Pi_m$. 
The algorithm then builds the projectors one at a time, upon defining 
their blocks iteratively. 
As we will see soon, initially the blocks of the first projector are 
mostly zero, and the building 
of the following projectors populates other blocks. In this sense, 
the amount of non-zero rows 
and columns grows during the building of the projectors, and the 
size of the necessary auxiliary 
Hilbert space $\mathcal{H}_A$ is obtained only at the end of the 
procedure.
\par
The algorithm initially builds the blocks of $E_1$ in order 
to satisfy the constraints \eqref{vincoliEm} on 
itself, i.e.
\beq
E_1 \cdot E_1 = E_1, \quad E_1 = E_1^\dagger, \quad E_1 \geq 0
\eeq
Then, we build \emph{some blocks} of $E_2$ in order to satisfy 
the orthogonality with $E_1$,
\beq
E_1 \cdot E_2 = 0
\eeq
and then imposing the other constraints 
\beq
E_2 \cdot E_2 = E_2, \quad  E_2 = E_2^\dagger,\quad  E_2 \geq 0
\eeq
we define the remaining  blocks. As we will see, this second step do not modify the previously 
defined blocks of $E_2$.
\par
Analogously, the algorithm builds $E_3$ (if any) imposing its orthogonality with $E_1,\ E_2$,  and then imposing  
that $E_3 \cdot E_3 = E_3$. The generalisation is straightforward, the element $E_m$ is built in order to satisfy 
at first the ortogonality with the previously built projectors, and then imposing the condition $E_m\cdot E_m = E_m$. 
The algorithm is thus an iterative one, since it employs the projectors already found, until all the elements are built.
\par
The algorithm requires basically two steps repeated several times: building a matrix with some assigned blocks
such that it is orthogonal to another matrix, and the completion of the matrix in order to make it idempotent, that is, 
satisfying 
$$
E_m \cdot E_m = E_m.
$$
These steps are analysed in some details in the following two Sections, whereas the overall algorithm
is summarised in Section \ref{ss:algo}.
\subsection{Building an idempotent matrix}
\label{costruzioneProiettore}
At first, let us consider the problem of building an idempotent matrix when some of its blocks are assigned. 
This is the case of the evaluation of $E_1$, which has the block $\Pi_1$ in the upper-left position. If $\Pi_1$ is already 
idempotent, we can just put $\Pi_1$ in the corner and set the remaining blocks to zero. If this is not the case, we can 
define the blocks around $\Pi_1$ such that $E_1 \cdot E_1 = E_1$, possibly employing the minimum amounts of 
blocks, and setting the others to zero. In what follows, we ignore the subscripts that refers to the $m$-th element. 
The general problem becomes to find the adjacent blocks of the upper-left corner in order to make the matrix 
$E$ idempotent. 
\par
As we will see in the following, it is enough to assume the following matrix structure for $E$
\beq
E = \left(\begin{matrix}
\Pi & A & 0 & \dots \\
A^\dagger & B & 0 & \dots \\
0 & 0 & 0 & \dots \\
\vdots & \vdots & \vdots & \ddots \\
\end{matrix}\right)
\eeq
with $\Pi$ a given block, while $A,\ B$ are blocks to find ($A^\dagger$ and $B\geq 0$ have been used so that $E = E^\dagger$).
In the case $\Pi^2 = \Pi$ we can omit $A,\ B$ since the matrix is already idempotent. Otherwise, we have to add the blocks $A,\ A^\dagger,\ B$ and the matrix 
$E$ grows in sizes. The constraint
\beq
E \cdot E = \left(\begin{matrix}
\Pi & A \\
A^\dagger & B  \\
\end{matrix}\right)
\cdot
\left(\begin{matrix}
\Pi & A \\
A^\dagger & B  \\
\end{matrix}\right)
=\left(\begin{matrix}
\Pi & A \\
A^\dagger & B  \\
\end{matrix}\right)
=E
\eeq
gives the following equations:
\begin{align}
\Pi^2 + A A^\dagger = \Pi \label{eq1} \\
\Pi A + A B = A \label{eq2} \\
A^\dagger A + B^2 = B \label{eq3}
\end{align}
Equation \eqref{eq1} can be solved exploiting the singular value decomposition (SVD) for 
$\Pi = V \Lambda V^\dagger$ and $A=USW^\dagger$. 
Setting $U=V,\ W=I,\ S=\sqrt{\Lambda (I-\Lambda)}$ leads to  $A=V\sqrt{\Lambda (I-\Lambda)}$. Assuming for the moment a full rank matrix $\Pi$, with eigenvalues strictly included in the range $(0,1)$, equation \eqref{eq2} allows us 
to find $B = I- \Lambda$.

Finally, equation \eqref{eq3} is verified with the above solutions,
and the blocks of $E$ can be built as
\beq
E = \left(\begin{matrix}
\Pi & A & 0 & \dots \\
A^\dagger & B & 0 & \dots \\
0 & 0 & 0 & \dots \\
\vdots & \vdots & \vdots & \ddots \\
\end{matrix}\right)
= \left(\begin{matrix}
V \Lambda V^\dagger & V \sqrt{\Lambda (I-\Lambda)} & 0 & \dots \\
\sqrt{\Lambda (I-\Lambda)}V^\dagger & I-\Lambda & 0 & \dots \\
0 & 0 & 0 & \dots \\
\vdots & \vdots & \vdots & \ddots \\
\end{matrix}\right)
\eeq
\par
A different route may be also employed upon exploiting positivity of the elements of the POVM. Indeed, for positive semi-definite
$\Pi$ we have the decomposition $\Pi~=~Y Y^\dagger$ (with $Y$ having no particular properties), which may be used instead 
of SVD, which is generally demanding in terms of computational time. Notice that if $\Pi$ is not full rank, the decomposition 
is still available, with $Y$ being a rectangular matrix with the same rank.

With this decomposition, equation \eqref{eq1} is solved by 
\beq
A=Y\sqrt{I - Y^\dagger Y}, \ A^\dagger = \sqrt{I - Y^\dagger Y} \  Y^\dagger. \label{solutionY}
\eeq
and $B = I- Y^\dagger Y$ follows. Finally, equation \eqref{eq3} is verified by re-writing $A^\dagger A = \sqrt{B} (I-B) \sqrt{B}$.

For rectangular $Y$, equation \eqref{solutionY} still holds upon defining $Y^{-1}$ as the Penrose inverse of Y, that is, 
the rectangular matrix satisfying $Y^{-1} Y = I$ on the support of $Y$.
In addition, the decomposition $E=Z Z^{\dagger}$ is also readily available from $Y$, 
\begin{align}
E =&  \begin{pmatrix}
YY^\dagger &  Y\sqrt{I - Y^\dagger Y} & 0 & \dots \\
\sqrt{I - Y^\dagger Y} \  Y^\dagger & I - Y^\dagger Y & 0 & \dots \\
0 & 0 & 0 & \dots \\
\vdots & \vdots & \vdots & \ddots \\
\end{pmatrix} 
\label{soluzioneCostruzioneProiettore}\\
& = 
\begin{pmatrix}
Y \\
\sqrt{I-Y^\dagger Y} \\
0 \\
\vdots
\end{pmatrix}
\cdot
\begin{pmatrix}
Y^\dagger & \sqrt{I-Y^\dagger Y} & 0 & \ldots 
\end{pmatrix}
= Z Z^{\dagger}
\label{scomposizioneIdempotente}
\end{align}

\subsection{Building a matrix orthogonal to a given one}
\label{costruzioneOrtogonale}
In this section we consider the problem of building a matrix (with some assigned blocks) 
such that it is orthogonal to a given one. This occurs in building, e.g., $E_2$, which has the 
upper-left block equal to $\Pi_2$ and must verify $E_1 \cdot E_2 = 0$. If we have $\Pi_1 \cdot \Pi_2 = 0$, 
it is enough to set the blocks adjacent to $\Pi_2$ equal to zero. In the most general case, this does not hold, 
and to satisfy the orthogonality condition we have to explicitly determine the blocks around 
$\Pi_2$. The expression
\beq
E_1 \cdot E_2 = \left(\begin{matrix}
Y_1Y_1^\dagger & Y_1 \sqrt{I-Y_1^\dagger Y_1} \\
\sqrt{I-Y_1^\dagger Y_1}\ Y_1^\dagger & I-Y_1^\dagger Y_1 \\
\end{matrix}\right)
\cdot
\left(\begin{matrix}
\Pi_2 & A \\
A^\dagger & B  \\
\end{matrix}\right)
=0\,,
\label{equazioneOrtogonale}
\eeq
where $\Pi_1 = Y_1 Y_1^\dagger$, provides the constraints
\begin{align}
Y_1Y_1^\dagger \Pi_2 + Y_1 \sqrt{I-Y_1^\dagger Y_1} A^\dagger=0 \label{eq4} \\
Y_1Y_1^\dagger A + Y_1 \sqrt{I-Y_1^\dagger Y_1} B=0 \label{eq5} \\
\sqrt{I-Y_1^\dagger Y_1} Y_1^\dagger A + (I-Y_1^\dagger Y_1) B = 0 \label{eq6}\,.
\end{align}
Equation \eqref{eq4} allows us to find $A = -\Pi_2 Y_1 \sqrt{I-Y_1^\dagger Y_1}$,
whereas equation \eqref{eq5} provides the expression  $B=\left(\sqrt{I-Y_1^\dagger Y_1}\right)^{-1} Y_1^\dagger \Pi_2 Y_1 \sqrt{I-Y_1^\dagger Y_1} $.
The third equation, \eqref{eq6}, is indeed verified by these solutions. 

At this stage, upon
imposing the orthogonality with $E_1$, we found that $E_2$ has the structure
\beq
E_2 = \left(\begin{matrix}
\Pi_2 & -\Pi_2 Y_1 \sqrt{I-Y_1^\dagger Y_1} & \quad \ast & \dots \\
- \left(\sqrt{I-Y_1^\dagger Y_1}\right)^{-1} Y_1^\dagger \Pi_2 &  \quad \left(\sqrt{I-Y_1^\dagger Y_1}\right)^{-1} Y_1^\dagger \Pi_2 Y_1 \sqrt{I-Y_1^\dagger Y_1} & \quad\ast & \dots \\
\ast & \ast & \quad\ast & \dots \\
\vdots & \vdots & \quad \vdots & \ddots \\
\end{matrix}\right),
\label{soluzioneCostruzioneOrtogonale}
\eeq
where the blocks indicated by $\ast$ are left unused and may be exploited to impose other conditions on $E_2$.
If a decomposition $\Pi_2=XX^{\dagger}$ is available, the big block just defined in \eqref{soluzioneCostruzioneOrtogonale} has a simple decomposition,  
\begin{align}
&\begin{pmatrix}
X X^\dagger & -X X^\dagger Y_1 \sqrt{I-Y_1^\dagger Y_1}  \\
- \left(\sqrt{I-Y_1^\dagger Y_1}\right)^{-1} Y_1^\dagger X X^\dagger &  \quad \left(\sqrt{I-Y_1^\dagger Y_1}\right)^{-1} Y_1^\dagger X X^\dagger Y_1 \sqrt{I-Y_1^\dagger Y_1} 
\end{pmatrix} \notag \\
& =\begin{pmatrix}
X \\
-\left(\sqrt{I-Y_1^\dagger Y_1}\right)^{-1} Y_1^\dagger X
\end{pmatrix}
\cdot
\begin{pmatrix}
X^\dagger &  \quad - X^\dagger Y_1 \sqrt{I-Y_1^\dagger Y_1}
\end{pmatrix}
=Y_2 \cdot Y_2^\dagger
\label{scomposizioneOrtogonale}
\end{align}

Notice that the blocks just defined depend on $\Pi_1$ (via its decomposition) and upon $\Pi_2$. If we have 
to impose the orthogonality of matrix $E_m$ with $E_1$, only the non-zero blocks in $E_1$ would be involved. Thus, the solution would be the same substituting $\Pi_m = X_m X_m^\dagger$ in place of $\Pi_2=X_2 X_2^\dagger$. 
\subsection{The algorithm}
\label{ss:algo}

The algorithm builds the projectors one at a time, using the previously built projectors. For each projector $E_m$ two steps are performed: first 
the {\em orthogonal construction} of Section \ref{costruzioneOrtogonale}, which defines some blocks of 
$E_m$ such that the projector is orthogonal to all the projectors previously evaluated. In the second step, leveraging the 
{\em idempotent construction} illustrated in Section \ref{costruzioneProiettore}, some other blocks are defined 
so that $E_m^2 = E_m$. Before applying the orthogonal or idempotent construction, it is checked whether $E_m$ is already ortogonal to the other projectors or idempotent. If this is the case, the step is simply skipped.
\par
The algorithm starts building $E_1$ with the idempotent construction, as the orthogonal one is not necessary. 
$\Pi_1$ is copied in the upper-left block of $E_1$ and 
the solution \eqref{soluzioneCostruzioneProiettore} is evaluated with $Y = Y_1,\ Y_1 Y_1^\dagger=\Pi_1$, where $Y_1$ has been obtained for instance from the singular value decomposition of $\Pi_1 = V_1 \Lambda_1 V_1^{\dagger}$, giving $ Y_1=V_1 \sqrt{\Lambda_1}$. If $\Pi_1$ is full rank, the 
projector $E_1$ has a nonzero $2\times2$ blocks in the upper-left corner. The remaining blocks are zero.
\begin{align*}
E_1 &= \left(\begin{array}{cc|cc}
      Y_1 Y_1^\dagger &  \quad Y_1 \sqrt{I-Y_1^\dagger Y_1} &       &  \\
      \ast &      I-Y_1^\dagger Y_1 & \multicolumn{2}{c}{\smash{\raisebox{.5\normalbaselineskip}{\quad 0}}} \\
			\hline \\[-\normalbaselineskip]
        &         &      &       \\
      \multicolumn{2}{c|}{\smash{\raisebox{.5\normalbaselineskip}{ 0}}}
			& \multicolumn{2}{c}{\smash{\raisebox{.5\normalbaselineskip}{\quad 0}}}
    \end{array}\right)
\end{align*}
The projector $E_2$ is then built using the two steps. The block $\Pi_2$ is copied in the upper-left corner, a 
decomposition $\Pi_2=X_2^{(1)} X_2^{(1)\dagger}$ is evaluated (by SVD if needed) and the three blocks around are 
defined as in \eqref{soluzioneCostruzioneOrtogonale} leveraging on the decompositon $\Pi_1 = Y_1Y_1^\dagger$ previously evaluated.
At this point, the big block just defined has a decomposition $Y_2 Y_2^{\dagger}$ as in \eqref{scomposizioneOrtogonale}, and if not idempotent, the adjacent blocks need to be evaluated accordingly employing the idempotent construction of equation \eqref{soluzioneCostruzioneProiettore}.
\begin{align*}
E_2 		
		& = \left( \begin{array}{c|c}
		\overbrace{
		\begin{matrix}
		\Pi_2 &  -\Pi_2 Y_1 \left(\sqrt{I-Y_1^\dagger Y_1}\right)^{-1}\\
		\ast &    \left(\sqrt{I-Y_1^\dagger Y_1}\right)^{-1} Y_1^\dagger  \Pi_2 Y_1 \left(\sqrt{I-Y_1^\dagger Y_1}\right)^{-1}
		\end{matrix}
		}^{Y_2 Y_2^\dagger}
		& \qquad Y_2 \sqrt{I-Y_2^\dagger Y_2}  \qquad \\
		\hline
	  \ast & \qquad  I-Y_2^\dagger Y_2  \qquad 
		\end{array} \right)
\end{align*}
Notice that in this case in the original matrix  $E_2$ the $4 \times 4$ blocks in the upper left corner are defined, 
for a total of  $4D$ rows and $4D$ columns (if  $\Pi_1,\ \Pi_2$ are full rank), with $D$ being the dimension of the system Hilbert space.
In the evaluation of $E_3$ the same orthogonal and idempotent construction are repeated, with the difference that the 
first must be repeated twice to get the orthogonality with $E_1$ ed $E_2$. As usual, first the block 
$\Pi_3 = X_3^{(1)} X_3^{(1)\dagger}$ is copied in the upper left corner. The first blocks around $\Pi_3$ are evaluated with \eqref{soluzioneCostruzioneOrtogonale}. 

The newly defined big block has decomposition $X_3^{(2)} X_3^{(2)\dagger}$ obtained from \eqref{scomposizioneOrtogonale} where $X=X_3^{(1)},\ Y_2 = X_3^{(2)}$. 
The orthogonal construction  \eqref{soluzioneCostruzioneOrtogonale} is repeated to get the orthogonality with $E_2$, employing the block $X_3^{(2)} X_3^{(2)\dagger}$ and the term $Y_2$ previously defined in the idempotent contruction of $E_2$. A new bigger block is 
obtained with decomposition $Y_3 Y_3^\dagger$ as in \eqref{scomposizioneOrtogonale}.

The idempotent construction is then used employing $Y = Y_3$ as in equation \eqref{soluzioneCostruzioneProiettore}.
Finally, we get the matrix (note that not all the blocks have the same size)
\begin{align*}
E_3 		
		= \left( \begin{array}{c|c}
		\overbrace{
		\begin{array}{c|c}
			\overbrace{
			\begin{matrix}
			\Pi_3 &  -\Pi_3 \ Y_1 R_1^{-1}\\
			\ast &    R_1^{-1} Y_1^\dagger \ \Pi_3 \ Y_1 R_1^{-1}
			\end{matrix}
			}^{X_3^{(2)} X_3^{(2)\dagger}}
			& \qquad -X_3^{(2)} X_3^{(2)\dagger} \ Y_2 R_2^{-1} \\
			\hline
			\ast & \displaystyle R_2^{-1} Y_2^\dagger \ X_3^{(2)} X_3^{(2)\dagger} \ Y_2 R_2^{-1}
		\end{array}
		}^{Y_3 Y_3^\dagger} & Y_3 R_3 \\
		\hline
		\ast & I-Y_3^\dagger Y_3
		\end{array} \right)
\end{align*}
with $R_1 = \sqrt{I-Y_1^\dagger Y_1},\ R_2 = \sqrt{I-Y_2^\dagger Y_2},\ R_3 = \sqrt{I-Y_3^\dagger Y_3} $.
Notice that the expression of $E_3$ depends upon the decompositions 
$Y_1Y_1^\dagger,\ Y_2Y_2^\dagger$ of the upper-left blocks of the 
preceding projectors. This holds for each projector $E_m$.
\par
The method used to evaluate  $E_3$ may be iterated for any subsequent projector. First, the block $\Pi_m = X_m^{(1)} X_m^{(1)\dagger}$ is 
copied in the upper left corner. The adjacent blocks are defined imposing the orthogonality with $E_1$,
 following the orthogonal construction. The just defined block has decomposition $X_m^{(2)} X_m^{(2)\dagger}$, and the 
 orthogonal construction is repeated using $Y_n$, which is the term used in the idempotent construction of 
 $E_n,\ n<m$. At the end of each orthogonal constructions, the newly defined big block is decomposed as $X_m^{(i)} X_m^{(i)\dagger},\ i<m$, and the orthogonal construction is repeated until $i$ reach $m$.
The term $X_m^{(m)} = Y_m$ is then used in the idempotent construction to get the final block structure of $E_m$. 
\par
At this stage, upon following the procedure leading to $E_m$, a recursive 
construction may be also obtained for its decomposition $E_m = Z_m Z_m^\dagger$. 
For further details, see \ref{a:cholesky}. 
If all the $\Pi_m$ are full rank and with eigenvalues in the range $[0,1]$, 
then the size of the projectors grows exponentially.  In fact, the projector 
$E_1$ has in this case $2\times2$ non-zero blocks, for a total of $2D$ rows 
and $2D$ columns; the projector  $E_2$ populates $4\times 4$ blocks, the 
projector  $E_3$ has $8 \times 8$ non-zero blocks, and so on.
An exception occurs if some of the blocks already satisfy the ortogonality 
conditions. For instance,  if $\Pi_2$ is already orthogonal to $\Pi_1$, 
there is no need to used the adjacent blocks to obtain 
its orthogonality. This is also the case if the block is idempotent, since 
the adjacent blocks may left unused.
\section{Examples}\label{s:exa}
Here we apply our procedure to obtain the Naimark extension of POVMs 
already presented in the literature. In this way, we are able to show 
the main features of the algorithm, and its advantages compared to
existing ones.
\subsection{Three elements POVM}
Helstrom considered the example a three-elements POVM 
$\{\Pi_1, \Pi_2, \Pi_3\}$, 
$\Pi_1+\Pi_2+\Pi_3 = {\mathbb I_S}$, 
defined by $\Pi_k = \frac23 \pure{\psi_k},\ 
k=1,2,3$, where \cite{hel76}
\beq
\ket{\psi_1}=\frac1{\sqrt{2}}
\left(\begin{array}{c}
1 \\ 1\end{array}\right)\,, \quad
\ket{\psi_2}=\frac1{\sqrt{2}}
\left(\begin{array}{c}
e^{-i \pi/3} \\
e^{i \pi/3} 
\end{array}\right)\,,\quad
\ket{\psi_3}=-\frac1{\sqrt{2}}
\left(\begin{array}{c}
e^{i \pi/3} \\
e^{-i \pi/3} 
\end{array}\right)\,.
\eeq
i.e.
\begin{align}
\Pi_1=& \frac13 \left(
\begin{array}{cc}
1 & 1 \\
1 & 1 \\
\end{array}
\right), \quad 
\Pi_2 = \frac13 \left(
\begin{array}{cc}
1& e^{- 2i \pi/3} \\
e^{2i \pi/3} & 1 \\
\end{array}
\right)\,, \quad
\Pi_3 = \frac13 \left(
\begin{array}{cc}
1& e^{2i \pi/3} \\
e^{-2i \pi/3} & 1 \\
\end{array}
\right)
\label{example3povm}
\end{align}
The extension originally obtained by Helstrom was based on a 
two-dimensional auxiliary Hilbert space with basis  
$\ket{v_1}=(1,0)^T$, $\ket{v_2}=(0,1)^T$,  and it is given by 
$E_k^H=\pure{\xi_k}$, $k=1,..,4$, where 
\begin{align}
& \ket{\xi_1} = \sqrt{2/3}\ket{v_1}\ket{\psi_1} 
+ \sqrt{1/3}\ket{v_2}\ket{\psi_3}, \\
& \ket{\xi_2} = \sqrt{2/3}\ket{v_1}\ket{\psi_2} 
- \sqrt{1/3}\ket{v_2}\ket{\psi_3}, \\
& \ket{\xi_3} = \sqrt{2/3}\ket{v_1}\ket{\psi_3} 
+ \sqrt{1/3}\ket{v_2}\ket{\psi_3}, \\
& \ket{\xi_4} = \ket{v_2}\ket{\psi_3'}, \\
& \ket{\psi_3'}=\frac1{\sqrt{2}}
\left(\begin{array}{c}
-e^{i \pi/3} \\
e^{-i \pi/3} 
\end{array}\right)\,.
\end{align}
The iterative algorithm in this case is particularly efficient since the orthogonality construction gives also 
idempotent matrices. Overall, a two-dimensional auxiliary Hilbert space is still required, but only the upper 
left 3-by-3 corner has non-zero coefficients.
\begin{align} \label{pvm}
& E_1 =  \frac{1}{3} \left(
\begin{array}{cccc}
1 & 1 & 1 & 0 \\ 
1 & 1 & 1 & 0 \\ 
1 & 1 & 1 & 0 \\ 
0 & 0 & 0 & 0 
\end{array}
\right), \quad
E_2 = \frac{1}{3} \left(
\begin{array}{cccc}
 1 & e^{-\frac{2 i \pi }{3}} & e^{\frac{2 i \pi }{3}} & 0 \\
 e^{\frac{2 i \pi }{3}} & 1 & e^{-\frac{2 i \pi }{3}} & 0 \\
 e^{-\frac{2 i \pi }{3}} & e^{\frac{2 i \pi }{3}} & 1 & 0 \\
 0 & 0 & 0 & 0 
\end{array}
\right), \\
& E_3 = \frac{1}{3} \left(
\begin{array}{cccc}
 1 & e^{\frac{2 i \pi }{3}} & e^{-\frac{2 i \pi }{3}} & 0 \\
 e^{-\frac{2 i \pi }{3}} & 1 & e^{\frac{2 i \pi }{3}} & 0 \\
 e^{\frac{2 i \pi }{3}} & e^{-\frac{2 i \pi }{3}} & 1 & 0 \\
  0 & 0 & 0 & 0 
\end{array}
\right)\,.
\end{align}
The correctness of both solutions is verified by checking the 
properties of orthogonality, idempotence, and the upper left 
corner equal to the original POVM. 
\par
The extension proposed by Helstrom gives 4-by-4 matrices with no zero coefficients, 
and therefore differs for the block adjacent the left upper corner. 
Here we report the matrix expression of $E_1^H$ for
comparison with $E_1$ in Eq. (\ref{pvm})
\begin{align}
& E_1^H= \frac{1}{3} \left(
\begin{array}{cccc}
 1 & 1 & \frac{e^{\frac{2 i \pi }{3}}}{\sqrt{2}} & \frac{e^{-\frac{2 i \pi }{3}}}{\sqrt{2}} \\
 1 & 1 & \frac{e^{\frac{2 i \pi }{3}}}{\sqrt{2}} & \frac{e^{-\frac{2 i \pi }{3}}}{\sqrt{2}} \\
 \frac{e^{-\frac{2 i \pi }{3}}}{\sqrt{2}} & \frac{e^{-\frac{2 i \pi }{3}}}{\sqrt{2}} & \frac{1}{2} &
   -\frac{1}{2} e^{-\frac{i \pi }{3}} \\
 \frac{e^{\frac{2 i \pi }{3}}}{\sqrt{2}} & \frac{e^{\frac{2 i \pi }{3}}}{\sqrt{2}} & -\frac{1}{2}
   e^{\frac{i \pi }{3}} & \frac{1}{2} \\
\end{array}
\right)\,.
\end{align}
\subsection{Four elements POVM}
Helstrom also considered a four-elements POVM 
$\{\Pi_1,\ \Pi_2,\ \Pi_3,\ \Pi_4\}$, with\cite{hel76}
$$
\Pi_k = \frac{1}{2}\pure{\psi_k}, \quad
  \ket{\psi_k} = \frac{1}{\sqrt{2}}
\left(\begin{array}{c}
e^{-i(k-1)\frac{\pi}{4}} \\
e^{i(k-1)\frac{\pi}{4}} 
\end{array} \right), \quad k= 1,2,3,4 \, ,
$$
i.e.
\beq
\Pi_1= \frac{1}{4}\left(
\begin{array}{cc}
 1 & 1 \\
 1 & 1 \\
\end{array}
\right),
\Pi_2=\frac{1}{4}\left(
\begin{array}{cc}
 1 & -i \\
 i & 1 \\
\end{array}
\right),
\Pi_3=\frac{1}{4}\left(
\begin{array}{cc}
 1 & -1 \\
 -1 & 1 \\
\end{array}
\right),
\Pi_4=\frac{1}{4}\left(
\begin{array}{cc}
 1 & i \\
 -i & 1 \\
\end{array}
\right).
\eeq
Again, the iterative algorithm easily finds the extension since the 
orthogonal construction directly gives idempotent matrices, without 
the need of the idempotent construction.
\begin{align}
&E_1=\frac{1}{4}\left(
\begin{array}{cccc}
 1 & 1 & \sqrt{2} & 0 \\
 1 & 1 & \sqrt{2} & 0 \\
 \sqrt{2} & \sqrt{2} & 2 & 0 \\
 0 & 0 & 0 & 0 \\
\end{array}
\right) 
&E_2=\frac{1}{4}\left(
\begin{array}{cccc}
 1 & -i & -e^{-\frac{i \pi }{4}} & e^{-\frac{i \pi }{4}} \\
 i & 1 & -e^{\frac{i \pi }{4}} & e^{\frac{i \pi }{4}} \\
 -e^{\frac{i \pi }{4}} & -e^{-\frac{i \pi }{4}} & 1 & -1 \\
 e^{\frac{i \pi }{4}} & e^{-\frac{i \pi }{4}} & -1 & 1 \\
\end{array}
\right) \\
&E_3=\frac{1}{4}\left(
\begin{array}{cccc}
 1 & -1 & 0 & i \sqrt{2} \\
 -1 & 1 & 0 & -i \sqrt{2} \\
 0 & 0 & 0 & 0 \\
 -i \sqrt{2} & i \sqrt{2} & 0 & 2 \\
\end{array}
\right) 
&E_4=\frac{1}{4}\left(
\begin{array}{cccc}
 1 & i & -e^{\frac{i \pi }{4}} & -e^{\frac{i \pi }{4}} \\
 -i & 1 & -e^{-\frac{i \pi }{4}} & e^{\frac{3 i \pi }{4}} \\
 -e^{-\frac{i \pi }{4}} & -e^{\frac{i \pi }{4}} & 1 & 1 \\
 e^{\frac{3 i \pi }{4}} & -e^{\frac{i \pi }{4}} & 1 & 1 \\
\end{array}
\right)
\end{align}

\subsection{Rank-2 POVMs}
In a more recent paper, rank-2 POVM elements have been 
introduced to describe generalized measurements involving 
sets of Pauli quantum observables chosen at random, the 
so-called {\em quantum roulettes} \cite{spar13}. More precisely,
quantum roulettes are generalized measurements obtained by selecting 
the observable $\sigma_k$  with a probability $\{z_k\}$ in the set of 
nondegenerate and isospectral observables $\{\sigma_k\}$. The 
POVM elements are defined as linear combination of the projectors 
associated with the observables outcomes.
\par
In Ref. \cite{spar13}, the \emph{canonical} Naimark extension is sought, 
i.e. the implementation of the generalized measurement in a larger 
Hilbert space using a projective indirect measurement on the ancillary 
system after its coupling with the system. In this scenario, 
Eq. \eqref{BornRule} is rewritten as
$$
\hbox{Tr}_A\big[\Pi_m\,\rho\big] = \hbox{Tr}_{AS}\big[ \left(\pure{\omega_A} \otimes \rho\right)\, U^\dagger (P_m \otimes {\mathbb I_S}) U\big]\,,
\label{BornRuleCanonical}
$$
where $\ket{\omega_A}$ is the ancillary state, $U$ describes the coupled 
evolution between the systems, and ${P_m}$ is the projective measurement 
in the ancillary system. A first example of POVM is that of a roulette 
obtained from the Pauli operators $\{\sigma_1, \sigma_3\}$ with 
probabilities $\{z, 1-z\},\ z\in(0,1)$, giving the elements 
$$
\Pi_1 = \frac{1}{2} \left(
\begin{array}{cc}
 2-z & z \\
 z & z \\
\end{array}
\right), \quad
\Pi_{-1} = \frac{1}{2} \left(
\begin{array}{cc}
 z & -z \\
 -z & 2-z \\
\end{array}
\right).
$$
The solution proposed uses the ancillary state $\ket{\omega_A} = \frac{1}{\sqrt{2}}\left(\ket{0} + e^{i\phi} \ket{1}\right)$, the projectors 
$$
P_1=\frac{1}{2} \left(
\begin{array}{cc}
 2-z & \sqrt{z(2-z)} \\
 \sqrt{z(2-z)} & z \\
\end{array}
\right), \quad
P_{-1}={\mathbb I} - P_1, 
$$
and the unitary 
$$
U=\left(
\begin{array}{cccc}
f & 0 & 0 & 0\\
0 & 0 & if^\ast & 0 \\
0 & if^\ast & 0 & 0 \\
0 & 0 & 0 & f 
\end{array}
\right), \quad f=\sqrt{\sqrt{\frac{2-2z}{2-z}} + i\sqrt{\frac{z}{2-z}}} \, .
$$
On the other hand, upon 
applying the iterative algorithm gives these solutions straightaway, 
$$
E_1=\left(
\begin{array}{cccc}
 1-\frac{z}{2} & \frac{z}{2} & \frac{\sqrt{(1-z) z}}{\sqrt{2}} &
   0 \\
 \frac{z}{2} & \frac{z}{2} & 0 & \frac{\sqrt{(1-z) z}}{\sqrt{2}}
   \\
 \frac{\sqrt{(1-z) z}}{\sqrt{2}} & 0 & \frac{z}{2} & -\frac{z}{2}
   \\
 0 & \frac{\sqrt{(1-z) z}}{\sqrt{2}} & -\frac{z}{2} &
   1-\frac{z}{2} \\
\end{array}
\right), \quad 
E_{-1} = {\mathbb I_{AS}} - E_1,
$$
which is equivalent to the canonical one up to a rotation in 
the ancillary state.

The paper presents also another example with rank-2 diagonal POVM elements,
$$
\Pi_1 = \left(
\begin{array}{cc}
\frac{1}{2}+ f & 0 \\
0 & \frac{1}{2}-f
\end{array}
\right), \quad
\Pi_{-1} = {\mathbb I} - \Pi_1.
$$
The proposed extension employs the ancillary state $\ket{\omega_A}=
\ket{e_1}$, the projectors of the observable $\sigma_3$, 
i.e. $P_1=\pure{e_1},\ P_{-1}= \pure{e_2}$, and the unitary
$$
U=\left(
\begin{array}{cccc}
 \sqrt{\frac{1}{2}+f} & 0 & 0 & i\sqrt{\frac{1}{2}-f} \\
 0 & \sqrt{\frac{1}{2}-f} & i \sqrt{\frac{1}{2}+f} & 0 \\
 0 & i \sqrt{\frac{1}{2}+f} & \sqrt{\frac{1}{2}-f} & 0 \\
 i\sqrt{\frac{1}{2}-f} & 0 & 0 &  \sqrt{\frac{1}{2}+f} \\
\end{array}
\right), 
$$
which gives
\beq
U^\dagger \ (P_1 \otimes {\mathbb I_S}) U = \left(
\begin{array}{cccc}
 \frac{1}{2}+f & 0 & 0 & \frac{1}{2} i \sqrt{1-4 f^2} \\
 0 & \frac{1}{2}-f & \frac{1}{2} i \sqrt{1-4 f^2} & 0 \\
 0 & -\frac{1}{2} i \sqrt{1-4 f^2} &  \frac{1}{2}+f & 0 \\
 -\frac{1}{2} i \sqrt{1-4 f^2} & 0 & 0 & \frac{1}{2}-f \\
\end{array}
\right).
\label{Can2}
\eeq
In this case the iterative algorithm is particularly easy to 
apply since we have diagonal POVM elements, and it gives the 
solution
$$
E_1 = \left(
\begin{array}{cccc}
 \frac{1}{2}+f & 0 & \frac{1}{2} \sqrt{1-4 f^2} & 0 \\
 0 & \frac{1}{2}-f & 0 & \frac{1}{2} \sqrt{1-4 f^2} \\
 \frac{1}{2} \sqrt{1-4 f^2} & 0 & \frac{1}{2}-f & 0 \\
 0 & \frac{1}{2} \sqrt{1-4 f^2} & 0 &  \frac{1}{2}+f \\
\end{array}
\right)
$$
which is equivalent to \eqref{Can2} since in both cases 
we can see $\Pi_1$ in the upper left bock.
\section{Conclusions}
\label{s:outro}
In this paper we have addressed the problem of finding the Naimark extension of a probability operator-valued 
measure, i.e. its implementation as a projective measurement in a larger Hilbert space. As a matter of fact, 
the extension of a POVM is not unique and we have exploited this degree of freedom to introduce an iterative 
method to build the projective measurement from the sole requirements of orthogonality and positivity. 
Our method improves existing ones, as  it is more effective in terms of computational steps needed to determine
the POVM extension. Even more importantly, our method may be employed also to extend POVMs containing 
elements with rank larger than one. 
\par
Since a Naimark extension provides a concrete model to realize the generalized measurement, we foresee applications 
of our method to assess technological solutions on different platforms and  to investigate the tradeoff between information 
gain and measurement disturbance in generalized measurements.
\section*{Acknowledgments}
This work has been supported by EU through the
Collaborative Project QuProCS (Grant Agreement 641277) and by UniMI
through the H2020 Transition Grant 15-6-3008000-625.
\appendix
\section{Kronecker product convention}
\label{a:kronecker}
The product space is usually defined as $\mathcal{H}_S \otimes \mathcal{H}_A$, with
the system Hilbert space $\mathcal{H}_S$ on the left.  However, given the definition of 
Kronecker product
$$
A\otimes B = \left(\begin{matrix} a_{11} B & \cdots & a_{1n}B \\ \vdots & \ddots & \vdots \\ a_{m1} B 
& \cdots & a_{mn} B \end{matrix}\right)\,,
$$
the opposite convention, i.e. describing the composite system by the Hilbert space 
$\mathcal{H}_A \otimes \mathcal{H}_S$, makes it easier to graphically visualize the 
product matrix. For instance, for a matrix given by the product of
the first element of the canonical basis only one block is non-zero
$$
(e_1 \cdot e_1^T)\otimes B  = 
\left(\begin{matrix}  B & 0 & \cdots & 0 \\ 0 & 0 & \cdots & 0 \\ \vdots & \vdots &\ddots & \vdots \\ 0 & 0 & \cdots & 0 
\end{matrix}\right)\,,
$$
The standard convention would make the notation more cumbersome.
\section{Building the decomposition of $E_m$}
\label{a:cholesky}
The procedure explained in Section \ref{ss:algo} suggests a recursive 
construction to directly obtain the decomposition $E_m = 
Z_m Z_m^\dagger$. In order to evaluate $Z_m$, we initially need a 
 decomposition $\Pi_m = X_m^{(1)}X_m^{(1)\dagger}$, 
obtained for instance from its singular value decomposition. Then, 
the orthogonal construction \eqref{scomposizioneOrtogonale} is 
applied with $X = X_m^{(i)},\ Y_1=Y_i$ to evaluate $ Y_2 = X_m^{(i+1)}$. 
This step is repeated for $i=1$ to $m$. 
The last block calculated, $X_m^{(m)}$, is defined as $Y_m$ and used 
in the idempotent construction \eqref{scomposizioneIdempotente} 
employing $Y=Y_m$ to get $Z = Z_m$.

This construction can be summarized by the following matrix (in general rectangular)
\beq
Z_m =\left(
\begin{matrix}
	\left . \begin{matrix}
		\left . \begin{matrix}
			\left . \begin{matrix}
				\left . \begin{matrix}
				X_m^{(1)} \\
				-\left(\sqrt{I-Y_1^\dagger Y_1}\right)^{-1} Y_1^\dagger X_m^{(1)}
				\end{matrix}  \right \} X_m^{(2)} \\
				\\
				-\left(\sqrt{I-Y_2^\dagger Y_2}\right)^{-1} Y_{2}^\dagger X_m^{(2)}
			\end{matrix}  \right \} X_m^{(i)} \\
			\\
			-\left(\sqrt{I-Y_i^\dagger Y_i}\right)^{-1} Y_i^\dagger X_m^{(i)}
		\end{matrix}  \right \} X_m^{(i+1)} \\
		\vdots \\
		-\left(\sqrt{I-Y_{m-1}^\dagger Y_{m-1}}\right)^{-1} Y_{m-1}^\dagger X_m^{(m-1)}
	\end{matrix}  \right \} X_m^{(m)}=Y_m \\
	\\
	\sqrt{I-Y_m^\dagger Y_m}
\end{matrix}\right)
\label{tYm}
\eeq
Notice that to obtain the term $Z_m$ the decomposition $X_m$ of $\Pi_m$ 
is used, as well as all the terms $Y_1,\ Y_2,\ \dots Y_{m-1}$ used in the preceding idempotent constructions. This is an efficient procedure, since 
the terms such as $\left(\sqrt{I-Y_{i}^\dagger Y_{i}}\right)^{-1} 
Y_{i}^\dagger,\ i<m$ are used in the later evaluation of the projectors, 
without the need to evaluate them at each iteration. 
Notice that also in this procedure we should check whether the matrices 
$X_m^{(i)} X_m^{(i)\dagger}$ are orthogonal to $E_i$ or if $Y_m 
Y_m^\dagger$ is idempotent. In this cases, there is no need to perform 
the orthogonal or idempotent construction of the algorithm.


\begin{thebibliography}{99}
\bibitem{nmk40} M. A. Naimark, Iza. Akad. Nauk USSR, Ser. Mat. {\bf 4} 277
(1940); C.R. Acad. Sci. URSS {\bf 41}, 359, (1943).
\bibitem{akh63} N. I. Akhiezer, I. M. Glazman, {\em Theory of Linear 
Operators in Hilbert Space} (Ungar, New York, 1963), Vol. 2.
\bibitem{hel73} C. W. Helstrom, Int. J. Theor. Phys. {\bf 8}, 361, (1973).
\bibitem{hel76} C. W. Helstrom, {\em Quantum Detection and Estimation
Theory} (Academic Press, New York, 1976)
\bibitem{hol01} A. S. Holevo, {\em Statistical Structure of
Quantum Theory}, Lect. Not. Phys {\bf 61}, (Springer, Berlin,2001).
\bibitem{per90} A. Peres, Found. Phys. {\bf 20}, 1441 (1990).
\bibitem{bin07} B. He, J. A. Bergou, Z. Wang, Phys. Rev. A {\bf 76}, 042326 (2007).
\bibitem{ber10} J. Bergou, J. Mod. Opt. {\bf 57}, 160 (2010).
\bibitem{mtqm} M. G. A. Paris, Eur. Phys. J. ST {\bf 203}, 61 (2012).
\bibitem{ban97} M. Ban, Int. J. Theor. Phys. {\bf 36}, 2583 (1997).
\bibitem{Zmeas}{M. G. A. Paris, G. Landolfi, G. Soliani}, 
J. Phys. A {\bf 40}, F531 (2007).
\bibitem{ecos03} R. Jozsa, M. Koashi, N. Linden, S. Popescu, S. Presnell, D. Shepherd, A. Winter, Quantum Inf. Comp. {\bf 3}, 405 (2003).
\bibitem{tam1} D. De Falco, D. Tamascelli, RAIRO Th. Inf. Appl.  
{\bf 40}, 93 (2006).
\bibitem{ben07} R. Beneduci, J. Math. Phys. {\bf 48}, 022102 (2007).
\bibitem{ben10} R. Beneduci, Int. J. Th. Phys. {\bf 49}, 3030 (2010).
\bibitem{lev89} R. Y. Levine, R. R. Tucci, Found. Phys. {\bf 19}, 175 (1989).
\bibitem{spar14} C. Sparaciari,  M. G. A.  Paris, Int. J. Quantum Inf. 
{\bf 12}, 1461012 (2014).
\bibitem{fey16}
 D. Tamascelli, S. Olivares, C. Benedetti, M. G. A. Paris, 
 Phys. Rev. A {\bf 94}, 042129 (2016).
\bibitem{igmd1} K. Banaszek, Phys. Rev. Lett. {\bf 86}, 1366 (2001).
\bibitem{igmd2} K. Banaszek, Open Sys. Information Dyn. {\bf 13}, 1 (2006)
\bibitem{igmd3} M. G. Genoni, M. G. A. Paris, Phys. Rev. A {\bf 71}, 052307  (2005).
\bibitem{igmd4} L. Mi$\check{\mbox{s}}$ta, R. Filip, Phys. Rev. A {\bf 72}, 
034307 (2005).
\bibitem{igmd5} M. G. Genoni, M. G. A.  Paris, Phys. Rev. A {\bf 74}, 012301 (2006).
\bibitem{igmd6} M. G. Genoni, M. G. A.  Paris, J. Phys. CP {\bf 67}, 012029 (2007).
\bibitem{wil13} M. Wilde, Proc. Roy. Soc. A {\bf 469}, 20130259 (2013).
\bibitem{iqc} A. Mandilara, J. W. Clark, Phys. Rev. A {\bf 71}, 013406 (2005).
\bibitem{per95} A. Peres, {\em Quantum Theory: Concepts and Methods} (Kluwer Academic, New York,1995).
\bibitem{pre98} J. Preskill, {\em Lecture Notes for Physics 229: Quantum Information and Computation}, California Institute of Technology, 1998.
\bibitem{spar13} C. Sparaciari,  M. G. A.  Paris, Phys. Rev. A {\bf 87}, 
012106 (2013).
\end{thebibliography}
\end{document}